\documentclass{article}
\usepackage{todonotes}
\usepackage{bm}
\usepackage{graphicx}

\usepackage{floatrow}
\usepackage{subfig}
\usepackage[font=small]{caption}
\usepackage{siunitx}
\usepackage{amsmath}
\usepackage{appendix}

\def\bmx{\bm{x}}

\def\bmchi{\bm{\chi}}

\def\bfK{\mathbf{K}}

\usepackage{PRIMEarxiv}

\usepackage[utf8]{inputenc} 
\usepackage[T1]{fontenc}    
\usepackage{hyperref}       
\usepackage[capitalize]{cleveref}
\crefname{appsec}{Appendix}{Appendices} 

\usepackage{url}            
\usepackage{booktabs}       
\usepackage{amsfonts}       
\usepackage{nicefrac}       
\usepackage{microtype}      
\usepackage{lipsum}
\usepackage{fancyhdr}       
\usepackage{graphicx}       
\usepackage{setspace}       
\graphicspath{{media/}}     
\usepackage{lineno}

\title{Elucidating Interfacial Dynamics of Ti--Al Systems \\ Using Molecular Dynamics Simulation and \\ Markov State Modeling
}

\author{
  Tianjiao Li, Chenxi Tian, Atieh Moridi, Jingjie Yeo\thanks{Corresponding author. Email: jingjieyeo@cornell.edu} \\
  Sibley School of Mechanical and Aerospace Engineering \\
  Cornell University \\ 
  Ithaca, NY 14853, United States\\
}

\begin{document}
\maketitle

\begin{abstract}
Due to their remarkable mechanical and chemical properties, Ti--Al based materials are attracting considerable interest in numerous fields of engineering, such as automotive, aerospace, and defense. With their low density, high strength, and resistance to corrosion and oxidation, these intermetallic alloys and compound metal-metallic composites have found diverse applications. However, additive manufacturing and heat treatment of Ti–Al alloys frequently lead to brittleness and severe formation of defects. The present study delves into the interfacial dynamics of these Ti--Al systems, particularly focusing on the behavior of Ti and Al atoms in the presence of TiAl\textsubscript{3} grain boundaries under experimental heat treatment conditions. Using a combination of Molecular Dynamics and Markov State Model analyses, we scrutinize the kinetic processes involved in the formation of TiAl\textsubscript{3}. The Molecular Dynamics simulation indicates that at the early stage of heat treatment, the predominating process is the diffusion of Al atoms towards the Ti surface through the TiAl\textsubscript{3} grain boundaries. The Markov State Modeling identifies three distinct dynamic states of Al atoms within the Ti/Al mixture that forms during the process, each exhibiting a unique spatial distribution. Using transition timescales as a qualitative measure of the rapidness of the dynamics, it is observed that the Al dynamics is significantly less rapid near the Ti surface compared to the Al surface. Put together, the results offer a comprehensive understanding of the interfacial dynamics and reveals a three-stage diffusion mechanism. The process initiates with the premelting of Al, proceeds with the prevalent diffusion of Al atoms towards the Ti surface, and eventually ceases as the Ti concentration within the mixture progressively increases. The insights gained from this study could contribute significantly to the control and optimization of manufacturing processes for these high-performing Ti--Al based materials.
\end{abstract}


\section{Introduction}
Ti--Al based materials have attracted significant attention in engineering due to their remarkable mechanical and chemical properties. The intermetallic alloy titanium aluminide (TiAl), for instance, is highly sought after in the aviation and automotive industries for its low density, high strength, and immunity to corrosion and oxidation.\cite{emiraliouglu2022additive} In addition, composite materials such as Ti/TiAl\textsubscript{3} and Al/TiAl\textsubscript{3} compound metal-metallic composites, which are based on Ti and Al, have also gained interest in the aerospace and defense industries for their high yield strength at elevated temperatures, high energy absorption capability, dimensional stability, and corrosion/oxidation resistance.\cite{zhang2022formation,ma2020microstructure} 

Nevertheless, fabrication of Ti--Al based materials can sometimes be challenging. Ti--Al intermetallics such as TiAl and TiAl\textsubscript{3} are brittle at room temperature, making them difficult to produce by conventional methods such as machining or shaping.\cite{emiraliouglu2022additive,dilip2017novel} The production of Ti--Al based materials is often achieved through various methods such as powder metallurgy for alloys,\cite{emiraliouglu2022additive} rolling followed by heat treatment or hot pressing for laminated structures,\cite{zhang2022formation,ma2020microstructure} and additive manufacturing.\cite{emiraliouglu2022additive} One of the key characteristics that these methods have in common is the use of elevated temperatures at specific stages. Notably, higher temperatures allow for increased diffusion of atoms.\cite{thiyaneshwaran2018nucleation} The diffusion at the interface between Ti and Al play a crucial role in the formation of various microstructures, including the Kirkendall voids.\cite{thiyaneshwaran2018nucleation} Such voids are the result of the Kirkendall effect, which is the motion of the interface between two metals that occurs as a consequence of the difference in diffusion rates of the metal atoms.\cite{wang2013hollow,el2015kirkendall} Presence of these voids is not desired in cases where the mechanical strength and stability of the metal part is crucial.\cite{sun2020fracture}

\begin{figure}[htbp]
    \centering
    \includegraphics[width=0.75\textwidth]{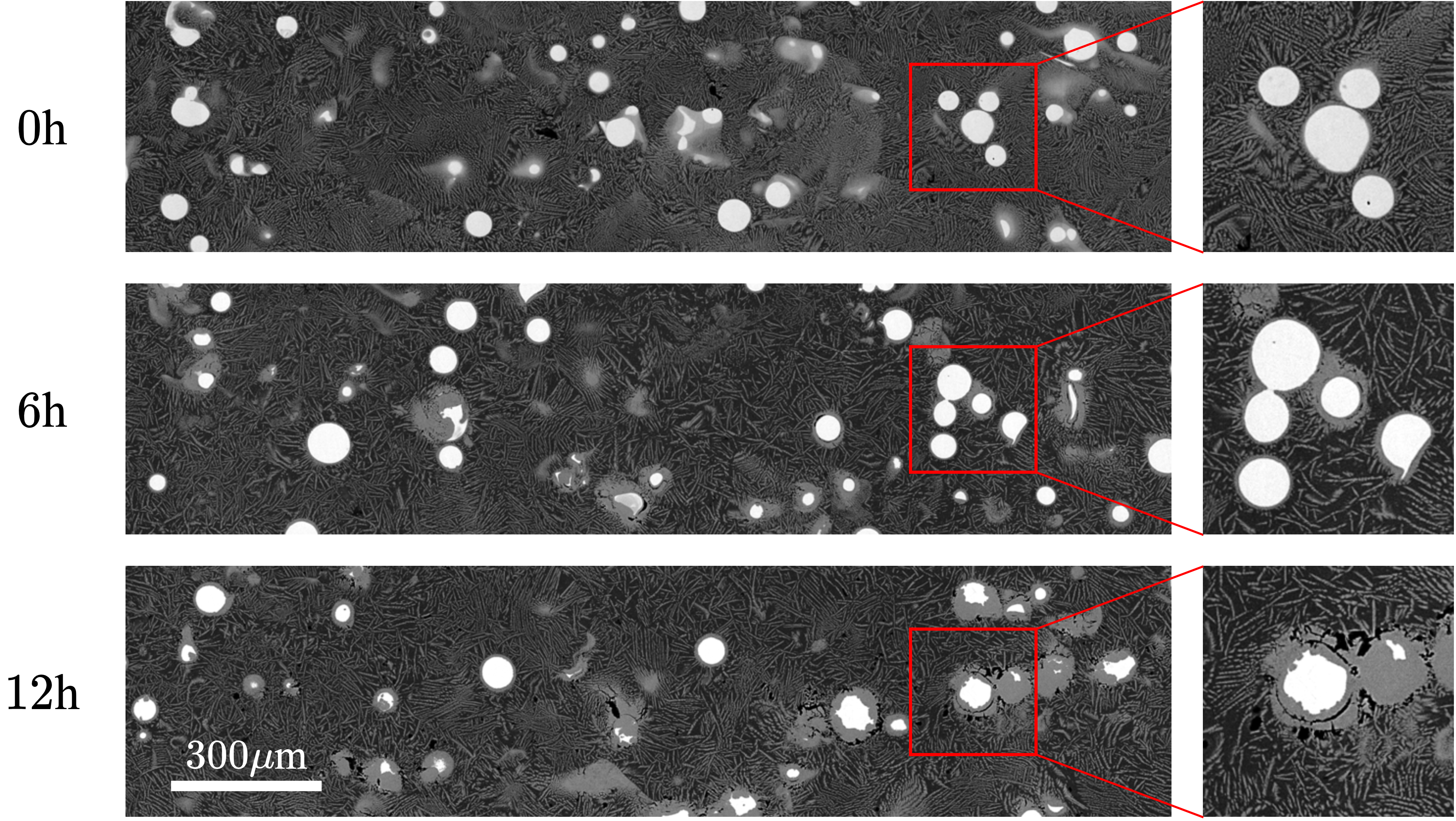}
    \caption{Cross-sectional views of a Ti--Al metal composite manufactured by additive manufacturing. The light regions in the image correspond to pure titanium, the gray regions correspond to TiAl\textsubscript{3} in the shape of dendrites that formed during manufacturing and layers that formed during heat treatment, and the dark regions correspond to pure aluminum. The images were taken using scanning electron microscope (SEM) and show the composite after 0 hours of heat treatment (top), 6 hours of heat treatment (middle), and 12 hours of heat treatment (bottom). To the right of each row, a zoom-in of a representative region of the corresponding view is shown to highlight the details of the microstructures near the Ti--Al interface.}
    \label{fig:kirkendall_experiment}
\end{figure}

There has been many experimental efforts to characterize the diffusion dynamics of Ti--Al binary systems. It is well established that TiAl\textsubscript{3} is the primary intermetallic phase that forms in the Ti/Al diffusion couple.\cite{mirjalili2013kinetics,zhang2022formation} A more recent study has revealed that in a Ti/Al binary diffusion couple, the diffusion of Al towards Ti is dominant, resulting in the concentrated nucleation of TiAl\textsubscript{3} at the Ti/TiAl\textsubscript{3} interface rather than the Al/TiAl\textsubscript{3} interface.\cite{thiyaneshwaran2018nucleation} However, current studies are primarily based on empirical relations such as the widely used model in which the thickness of reaction layer $x$ is linear in diffusion time $t$ raised to a kinetic exponent $n$, i.e. $x = Kt^n$, where $K$ is the rate constant.\cite{xu2006growth,mirjalili2013kinetics,zhang2022formation} 
These models are powerful tools when explaining and predicting regular systems such as laminated composites, but may not be easy to use when it comes to systems with complex interfaces and microstructures, such as those found in materials produced through additive manufacturing. 
\cref{fig:kirkendall_experiment} illustrates the cross-sectional views of one of our previous attempts to create Al/TiAl\textsubscript{3} composite using direct energy deposition (DED). The complex nature of the system is compounded by the presence of dendrites, dispersed pure Ti particles, and the non-uniform concentration gradient of Ti and Al. The images were captured using scanning electron microscope (SEM) and show the composite after 0 hours, 6 hours, and 12 hours of heat treatment. As the heat treatment time increases, the Kirkendall effect becomes more significant, as observed by the formation of voids around the Ti particles and intermetallics at the Ti/Al interface. These microstructural changes are a result of the difference in atomic diffusion rates between the titanium and aluminum components of the alloy, and are a clear demonstration of the Kirkendall effect in action. This highlights the importance of understanding the diffusion dynamics in the design and manufactruing of new alloys and composites. 

Molecular Dynamics (MD) simulations are a proven technique for understanding the atomic-level dynamics of materials, including alloys. Several previous studies have demonstrated the efficacy of MD in simulating the formation of Ti--Al intermetallics.\cite{kiselev2014molecular,kiselev2016modeling,levchenko2012molecular} Moreover, a recent MD study combined with experiments on the reactivity of the Ti--Al system have identified an exothermic atom substitution process at the Ti/Al interface, where liquid Al atoms replace solid Ti atoms, leading to continued dissolution from areas where multiple Ti atoms have been substituted in close proximity.\cite{bizot2020reactivity} Another recent study uses MD to investigate the self-diffusion in liquid and solid alloys of the Ti--Al system. The results show that the activation energy of self-diffusion depends on the concentration of components, and is smaller in disordered alloys than in ordered alloys.\cite{poletaev2021self} MD simulations are a powerful tool for understanding physical processes at the atomic level, including the vacancy diffusion mechanism.\cite{poletaev2021self} However, traditional MD analysis often relies on researchers observing and interpreting a few infrequent events in simulation movies. Such analysis can lead to a lack of statistical relevance for the events observed or to potentially miss crucial but rare events,\cite{prinz2011markov} such as surface diffusion at low temperatures and chemical processes with a high activation barrier.\cite{voter1985dynamical}

Markov State Models (MSM) provide a data-driven approach for analyzing molecular dynamics (MD) simulation data.\cite{hinrichs2007calculation,pande2010everything,husic2018markov} MSM partitions the dynamics of atoms or molecules into discrete states and calculates the transition probabilities based on either theory or data, thus producing a model of the dynamics that is free from the subjectivity of the analyst.\cite{prinz2011markov} Also, it is claimed that all valid equilibrium MD can be explained through the concept of relaxation processes that are represented by the eigenfunctions of the dynamical operator, making MSM a generalizable approach for MD data analysis.\cite{prinz2011markov} The development of the Variational Approach for Markov Processes (VAMP) offers a mathematical framework for optimizing Markov State Models (MSM) with time series data.\cite{wu2017variational,wu2020variational} In a follow-up work, the theory of VAMP is implemented in deep neural networks to learn the transition rate between molecular structures.\cite{mardt2018vampnets} Furthermore, the recently developed Graph Dynamical Networks (GDyNets) uses a graph convolutional neural network architecture that incorporates the graph representation of particle systems and the VAMP theory in learning atomic-scale dynamics in general phases and environments from MD simulations.\cite{xie2019graph} The machine-learned model, trained with a moderate trajectory sampling timespan, can capture the rare transitions in the material system, as demonstrated in the study.\cite{xie2019graph}

In our study, we first employ MD to scrutinize the behavior of Ti and Al atoms within TiAl\textsubscript{3} grain boundaries at elevated temperatures, thereby replicating the conditions of experimental heat treatment. Through our computational data, we explore the kinetics involved in the formation of TiAl\textsubscript{3} during this heat treatment process. Subsequently, we use GDyNets to build MSM and delve deeper into the dynamics of Al atoms in the Ti/Al amalgamation, taking into account the presence of Ti and Al substrates. By characterizing distinct dynamic states, mapping their spatial distributions, and determining the varied timescales associated with state transitions, we are able to provide a comprehensive analysis of the interfacial dynamics of Ti--Al systems, and therefore, shed light on the role of the dynamics in the formation of TiAl\textsubscript{3} during the manufacturing process.

\section{Method}
\subsection{Molecular simulation details}

\begin{figure}[htbp]
    \centering
    \includegraphics[width=0.6\textwidth]{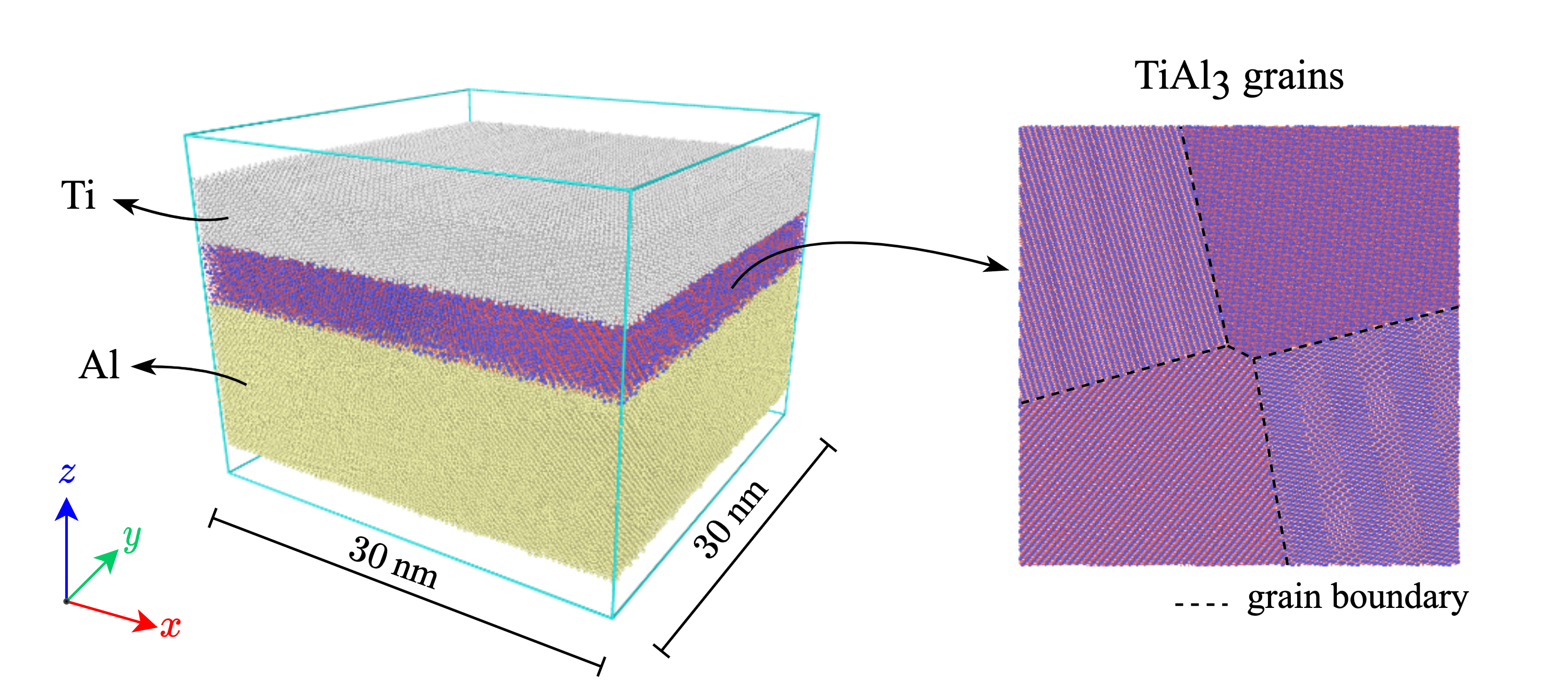}
    \caption{The simulation cell for the heat treatment process. Besides the dimensions shown in the figure, the thicknesses of the Al, TiAl\textsubscript{3} and Ti layers are \SI{10}{\nano\meter}, \SI{4}{\nano\meter} and \SI{4}{\nano\meter} respectively. The TiAl\textsubscript{3} layer consists of $4$ randomly oriented grains, forming grain boundaries. The lattice orientations of the Al and Ti layers are also random. 
    }
    \label{fig:heat_treatment}
\end{figure}

To investigate the diffusive dynamics of the atoms in realistic grains present in experimental settings,\cite{thiyaneshwaran2018nucleation,mirjalili2013kinetics} we define a simulation cell that resembles the interfacial structure of Ti--Al systems, as shown in \cref{fig:heat_treatment}.
The system has three layers, from top to bottom, there is a layer of pure Ti, a layer of TiAl\textsubscript{3} grains, and a layer of pure Al. The top \SI{1}{\nano\meter} layer of Ti and the bottom \SI{1}{\nano\meter} layer of Al are fixed in the xy-plane and cannot move freely. The z-axis movement is not restricted. A \SI{1}{\nano\meter}-wide margin of the TiAl\textsubscript{3} region is fixed in all three directions, preventing the TiAl\textsubscript{3} layer from floating freely in space.

The interatomic interactions in the Ti--Al system are modeled using the embedded atom model (EAM) potential developed in \cite{zope2003interatomic} by fitting to both experimental and \textit{ab initio} data. This potential model has proven to be effective in various investigations and have been successfully tested for many mechanical and structural-energy properties of Ti-Al systems.\cite{poletaev2021self,wu2016molecular,tang2014molecular,gao2023design} A recent comprehensive study has demonstrated that after over two decades since its creation, this potential model remains a highly effective representation of the physical properties of the Ti--Al systems, even compared to newer models.\cite{pei2021systematic,kim2016atomistic}

The system is first equilibrated at \SI{300}{\kelvin} for \SI{1}{\nano\meter}. Then, it is heated up to \SI{895.15}{\kelvin} in \SI{300}{\pico\second}, corresponding to a heating rate of approximately \SI{2}{\kelvin/\pico\second}. Next, the system is kept at \SI{895.15}{\kelvin} for \SI{30}{\nano\second}. \SI{895.15}{\kelvin} is a typical temperature used in \cite{mirjalili2013kinetics}. Besides, multiple experimental studies show a temperature range of \SI{825}{\kelvin} to \SI{1220}{\kelvin} for Ti/Al system heat treatment.\cite{bizot2020reactivity,mirjalili2013kinetics,xu2006growth,sun2011multilayered,lazurenko2016explosively}

\subsection{Markov State Modeling methods}
\subsubsection{Modeling dynamics with graph representation of atoms}
To describe the state transition of a dynamical system, there exists a state space $\Omega$ that contains all dynamical variables needed, and $\bmx_t \in \Omega$ is a dynamical process. To describe the dynamics of an atom in a MD simulation, we can think of $\Omega$ as the space of local configurations of the atom. Such local configurations can often be encoded by the species of the target atom, the species of the surrounding atoms, distances (or bond length) between target atom and its neighbors, and sometimes, even the angular components describing the angles formed by atoms with target atom being the vertex. For example, \cref{asec:atomic_environment} describes how a local configuration of an atom can be a represented by a weighted colored graph.


Then, the dynamics of the target atom can be described as a Markov process:\cite{xie2019graph}
\begin{equation}
    \bmx_{t+\tau} = \mathbf{F}(\bmx_t)
\end{equation}
where $\mathbf{F}$ is the dynamical system, which is non-linear in $\bmx$. Many recent studies have applied Koopman theory to build data-driven dynamical models. In short, Koopman theory states that there is a linearizing coordinate $\bmchi$ such that 
\begin{equation}
    \bmchi(\bmx_{t+\tau}) = \mathbf{K}(\tau)^\top\bmchi(\bmx_t)
\end{equation}
for some lag time $\tau$. It is shown that $\bfK$ has eigenvalues $1 \equiv \lambda_0 > \lambda_1 > \lambda_2 > \cdots > -1$.\cite{bowman2013introduction} The associated eigenvectors $\mathbf{v}_i$ can be viewed as principal modes of the dynamics, and the principal modes with the largest eigenvalues are the primary modes of probability flow between the system's substates \cite{bowman2013introduction}. Notably, $\mathbf{v}_0$ corresponds to the stationary distribution of the states, because raising $\lambda_0$ to any power does not change the weight of the process.\cite{prinz2011markov} For $i \neq 0$, the relaxation timescale of the $i$th slowest process is computed as
\begin{equation}
    t_i(\tau) = - \frac{\tau}{\ln \lvert \lambda_i(\tau) \rvert}
\end{equation}






In this work we use GDyNets developed in \cite{xie2019graph} to learn the Markov state model of the dynamics of the Ti-Al binary system. GDyNets is an implementation of the VAMP algorithm in TensorFlow dedicated to study atomistic level dynamics. The structure of the network and hyperparameter settings are described in \cref{asec:neural_network_architecture}.

\subsubsection{Sampling strategy}
\begin{figure}[htbp]
    \sidesubfloat[]{
        \includegraphics[height=8\baselineskip]{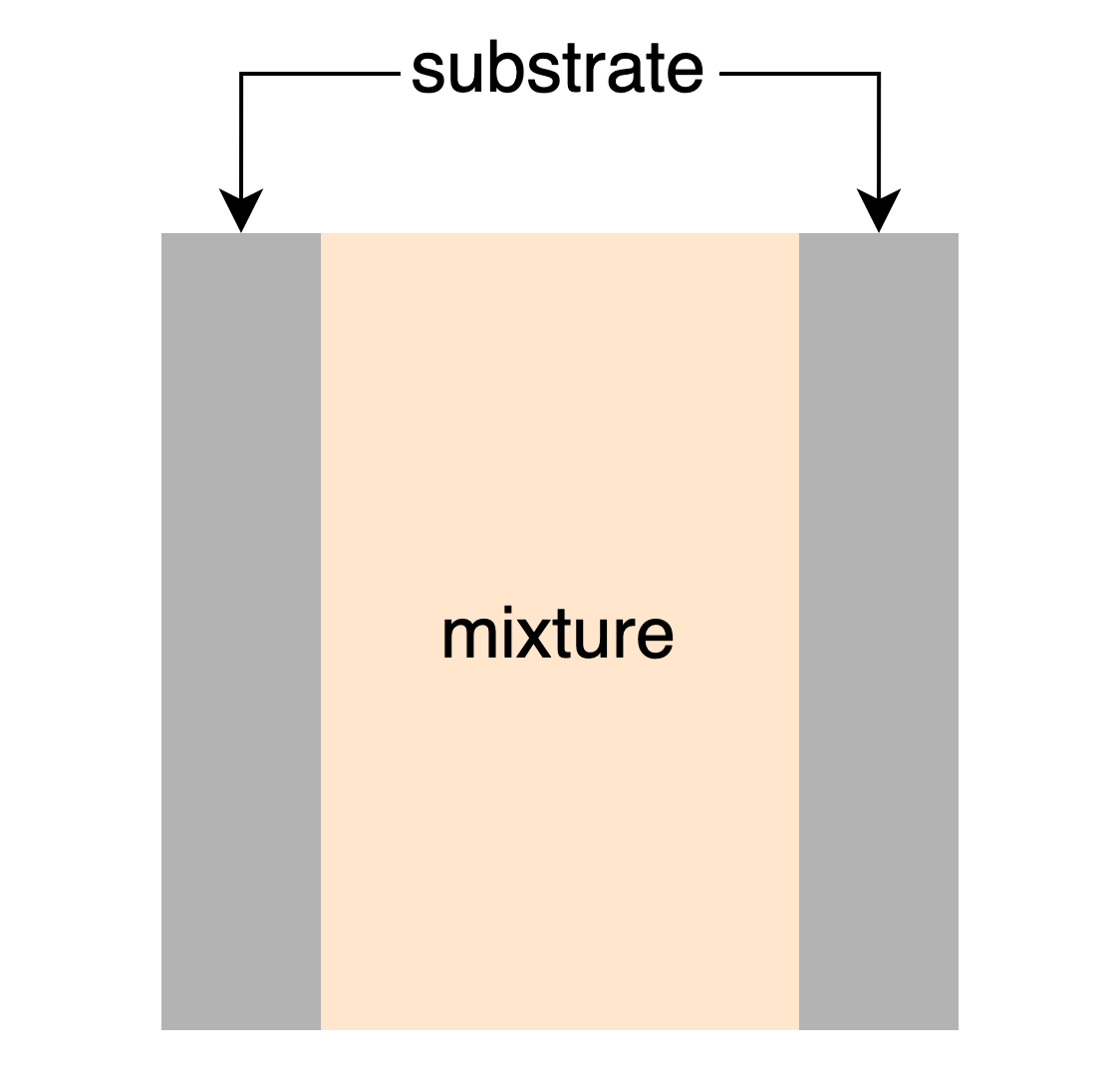}\label{fig:sandwich_struct}
        }
    \sidesubfloat[]{
        \includegraphics[height=8\baselineskip]{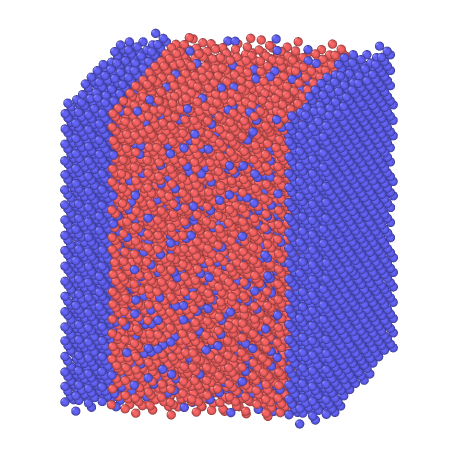}\label{fig:Ti-based-config}
        }
    \sidesubfloat[]{
        \includegraphics[height=8\baselineskip]{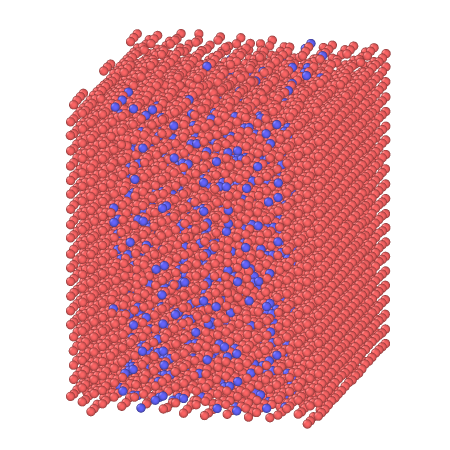}\label{fig:Al-based-config}
        }
    \caption{Schematics and examples of the sandwich configuration of the Ti-Al binary system. (a) Schematic of the sandwich structure; (b) The Ti-substrated configuration; (c) The Al-substrated configuration. The blue atoms represent the Ti atoms and the red atoms represent the Al atoms. A mixture of both species can be found in the middle of the sandwiches. In both configurations, the Ti-Al mixture consists of $7500$ atoms. The compositions of the mixtures are both $\mathrm{Ti}10\%\mathrm{Al}90\%$.}
    \label{fig:config}
\end{figure}

For the training, testing and validation purpose of the MSM, we sample the interfacial dynamics of the Ti--Al binary system using a sandwich structure as illustrated in Figure \cref{fig:config}. The sandwich structure is composed of a substrate of either Ti or Al, with the top and bottom four layers of atoms being made of Ti in Figure \cref{fig:Ti-based-config} or Al in Figure \cref{fig:Al-based-config}. The middle of the sandwich structure is comprised of a mixture of Ti and Al atoms, with a total of $7500$ atoms.

In our model, the substrates are treated as rigid bodies, while the $7500$-atom mixture is only subjected to thermostatting. Although in real experiments, no part of the system is truly rigid, this study places emphasis on the timescales of the transitions between dynamic states. The assumption of rigidity for the substrates is reasonable as the time required for the atoms in the solid substrate to transition into the liquid mixture is likely much longer than the other transitions occurring within the liquid mixture. Furthermore, modeling the substrates as rigid simplifies the system and ensures that it converges to a reversible dynamic equilibrium at a given temperature.

We focus on three key parameters of a sandwich structure: 1. the atomic species of the substrate $S = \{\mathrm{Ti}, \mathrm{Al}\}$, 2. the composition of the mixture $C = \{\mathrm{Ti}10\%\mathrm{Al}90\%, \mathrm{Ti}5\%\mathrm{Al}95\%, \mathrm{Ti}0\%\mathrm{Al}100\%\}$, and 3. the temperature $T = \{1033.5\mathrm{K}, 983.5\mathrm{K}\}$. Same as the temperature we set for the previous MD simulations, these two temperatures both fall in the range of typical Ti/Al system heat treatment. For an instance of the dynamical systems $s$, there is $s \in S \times C \times T$.

In the simulations, we use an overdamped Langevin thermostat, as suggested by \cite{prinz2011markov}. This ensures the dynamic process is Markovian, ergodic, and reversible, and thus satisfy the assumption of the theory of MSM developed in \cite{prinz2011markov}. For each instance of the dynamic system, we sample 5000 frames of the dynamics, with the time interval between every two frames being \SI{0.25}{\pico\second}. 2500 frames of the data are used for training, and 2000 for testing and 500 for validation.



\section{Results and Discussion}
\subsection{Molecular simulation of heat treatment}
\label{sec:md_results}
\begin{figure}[htbp]
    \def\subplotwidth{0.85\textwidth}
    \def\verticalskip{0.5\baselineskip}
    \newcommand{\subplot}[2]{
        \sidesubfloat[]{
            \includegraphics[width=\subplotwidth]{#1}\label{#2}
        }
    }
    \subplot{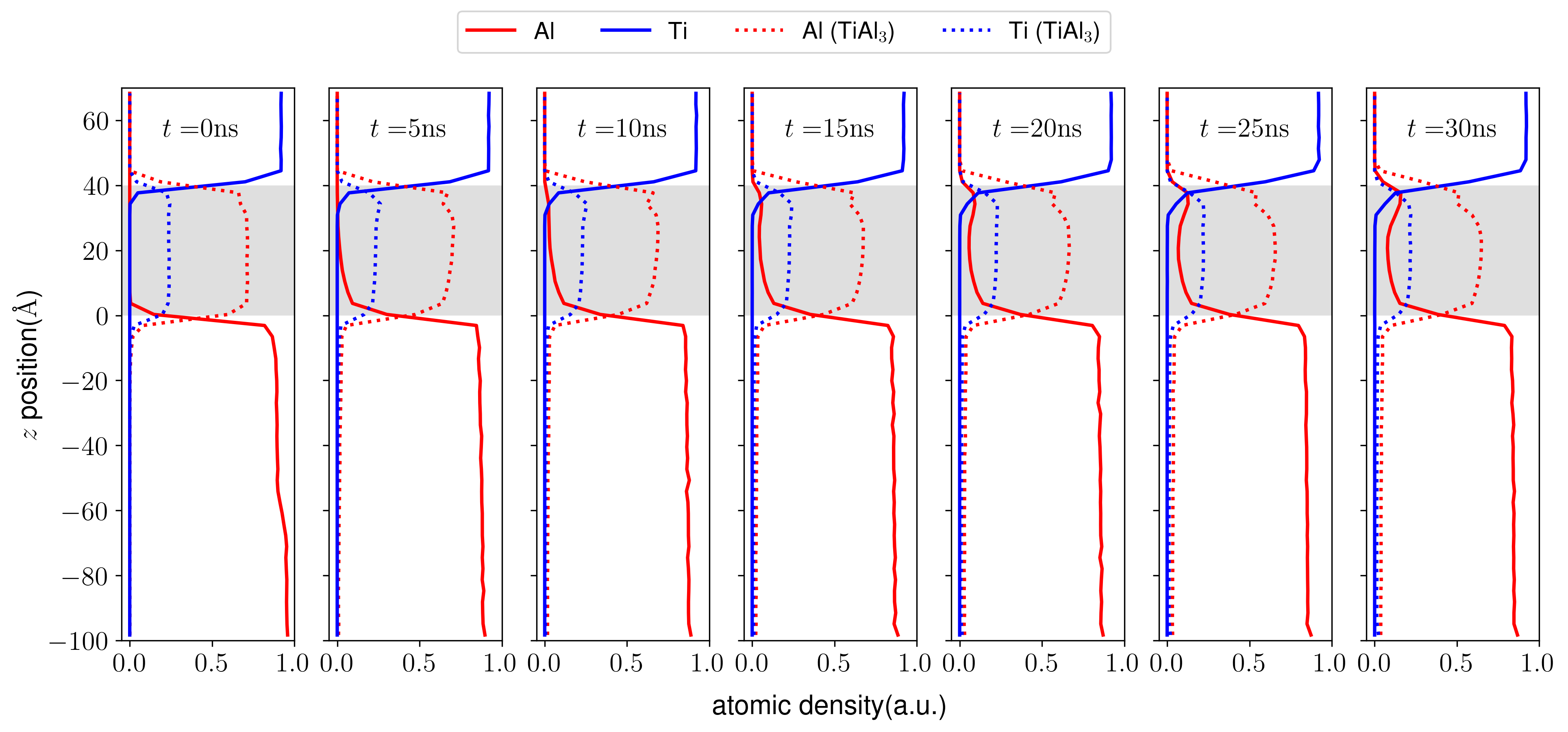}{fig:atomic_diffusion}
    \\
    [\verticalskip]
    \subplot{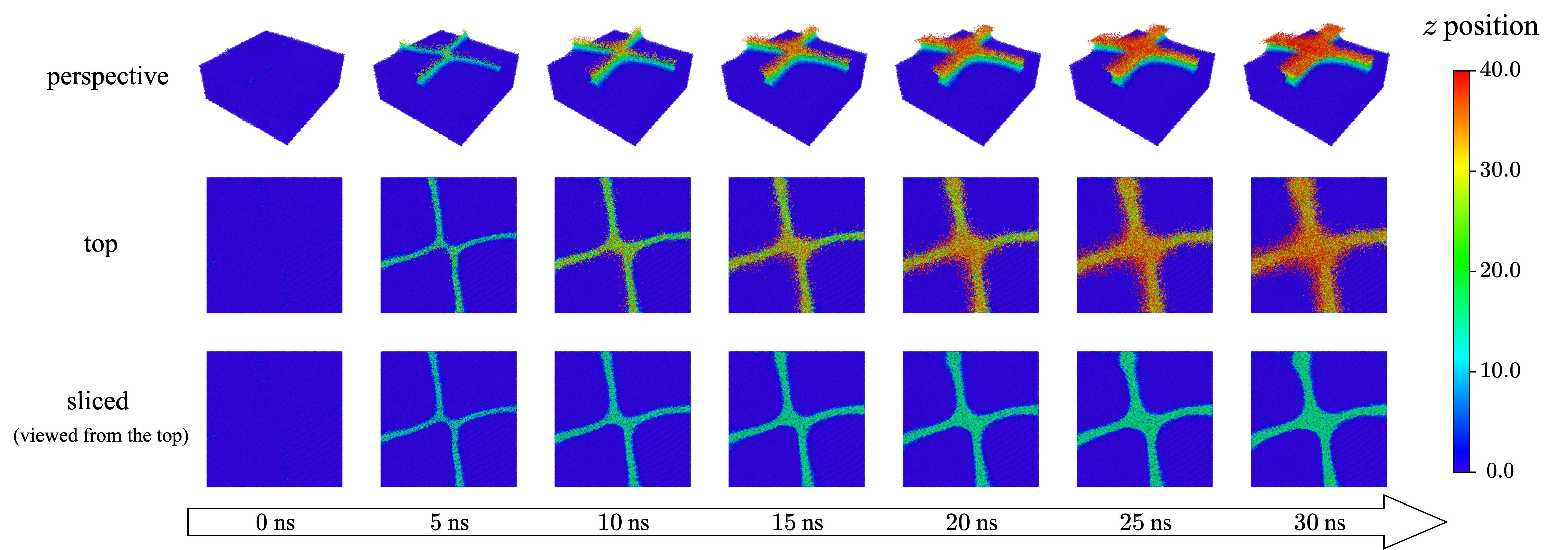}{fig:Al_diffusion}
    \caption{The dynamic evolution of atoms during heat treatment simulation. (a) shows the evolution of density profiles of all the atoms along the z-direction during heat treatment simulation. The shaded region corresponds to the TiAl\textsubscript{3} grain layer. (b) depicts the variation of the z-positions of Al atoms from the pure Al layer. The colorbar corresponds to the range of the TiAl\textsubscript{3} grain layer, with the lower limit representing the bottom and the upper limit representing the top of the layer. The entire pure Al region lies below the TiAl\textsubscript{3} layer and is shown in dark blue. The "sliced" visualizations in the bottom row present a top-view of the Al atoms with the atoms located above z-position $20$ removed, providing insight into the configuration of Al atoms in the TiAl\textsubscript{3} grain boundaries.}
    \label{fig:diffusion}
\end{figure}

The results of the MD simulation are summarized in \cref{fig:diffusion}. The simulated heat treatment was conducted at \SI{895.15}{\kelvin}, a temperature just below Al's melting point of \SI{933.5}{\kelvin} and significantly lower than Ti's melting point of \SI{1941}{\kelvin}. It should be noted that under these conditions, Al atoms are likely to exhibit more dynamic activity than Ti atoms. For Ti atoms in the pure Ti phase to disengage from their local environment and diffuse, a fundamental modification and activation of the Ti surface is necessary.

\cref{fig:atomic_diffusion} depicts the dynamic atom movements during the heat treatment process. It highlights a significant diffusion of Al atoms from the pure Al phase into the TiAl\textsubscript{3} grain boundaries ($0 < z < 40$ \AA), likely a result of pre-melting as evidenced by the observation of liquid Al at the interfaces. An apparent accumulation of Al atoms near the pure Ti phase is also observed, as indicated by an increased density near $z = 40$ \AA. However, the figure illustrates limited diffusion of Ti atoms, whether they initially reside in the pure Ti or the TiAl\textsubscript{3} phase. The Ti surface remains largely inactive in the simulation and doesn't exhibit substantial diffusion even when Al atoms infiltrate the Ti/TiAl\textsubscript{3} interface. Despite the co-existence of Al diffusion into the Ti side and Ti diffusion into the Al side, the former dominates. Ti atoms from the pure Ti phase rarely migrate into the Al side or into the TiAl\textsubscript{3} grain boundaries, suggesting that Ti atom diffusion might only be noticeable at very extended time-scales due to long activation times. Moreover, Al atoms initially in the TiAl\textsubscript{3} phase show negligible diffusion, likely due to the stability of the TiAl\textsubscript{3} lattice structure at the heat treatment temperature.

Therefore, the most predominant process is the diffusion of Al atoms from the pure Al layer towards the Ti layer via the TiAl\textsubscript{3} grain boundaries, followed by the accumulation of Al atoms at the Ti/TiAl\textsubscript{3} interface. \cref{fig:Al_diffusion} specifically focuses on this phenomenon. As demonstrated in the figure, during the heat treatment, the Al atoms diffuse along the TiAl\textsubscript{3} grain boundaries, causing them to slowly broaden. Upon reaching the Ti/TiAl\textsubscript{3} interface, the Al atoms exhibit a tendency to accumulate. This result is consistent with literature that the Ti-Al alloy forming mostly happens in the vicinity of pure Ti layer.\cite{thiyaneshwaran2018nucleation, mirjalili2013kinetics} The findings indicate that in additive manufacturing, post-processing a manufactured part with remaining pure Ti phase embedded in the Al matrix for full reaction of Ti with Al could be challenging. This difficulty arises due to the considerable disparity between the diffusion rates of Al and Ti, which may lead to the formation of Kirkendall voids at the Al/TiAl\textsubscript{3} interface. The issue can potentially be mitigated by the recently established \textit{in situ} reactive printing technique in additive manufacturing.\cite{tian2022situ} This method leverages on the rapid cyclic thermal conditions of metal additive manufacturing to instantaneously form reinforcing reaction product during the printing process.

\subsection{Distribution of states}
\label{sec:distribution_of_states}
The MD simulation result shows that the diffusion of pure Al atoms is dominant in this system. Therefore, we focus our analysis on the dynamic behavior of these atoms.

\begin{figure}[htbp]
    \def\subplotheight{0.18\textwidth}
    \def\verticalskip{0.5\baselineskip}
    \newcommand{\subplot}[2]{
        \sidesubfloat[]{
            \includegraphics[height=\subplotheight]{#1}\label{#2}
        }
    }

    \subplot{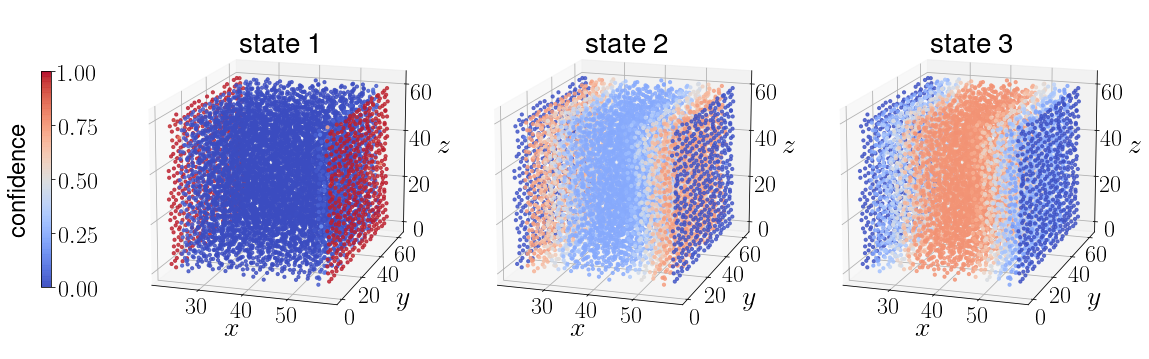}{fig:Al-distr_sub0}
    \subplot{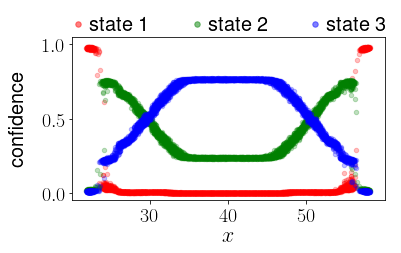}{fig:Al_prob_sub0}
    \\

    \subplot{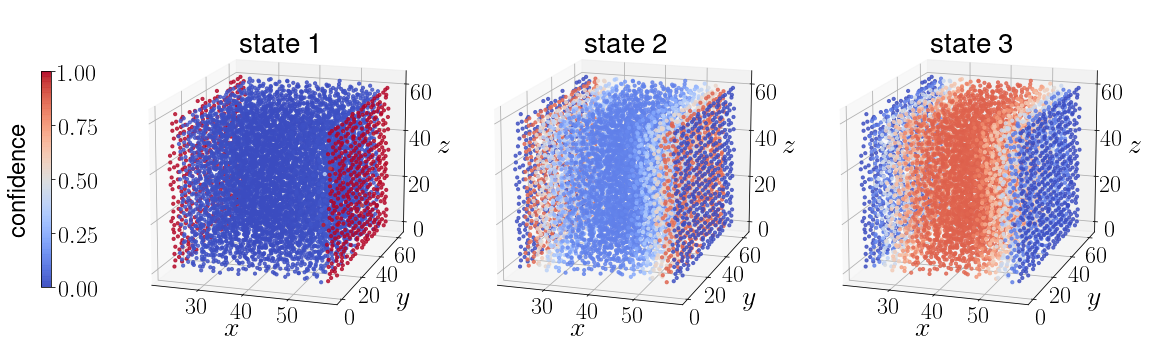}{fig:Al-distr_sub1}
    \subplot{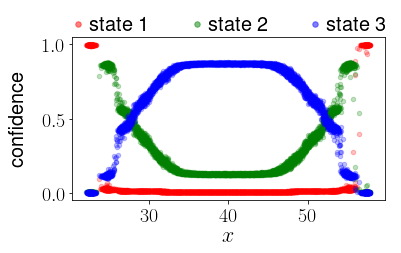}{fig:Al_prob_sub1}
    \\

    \subplot{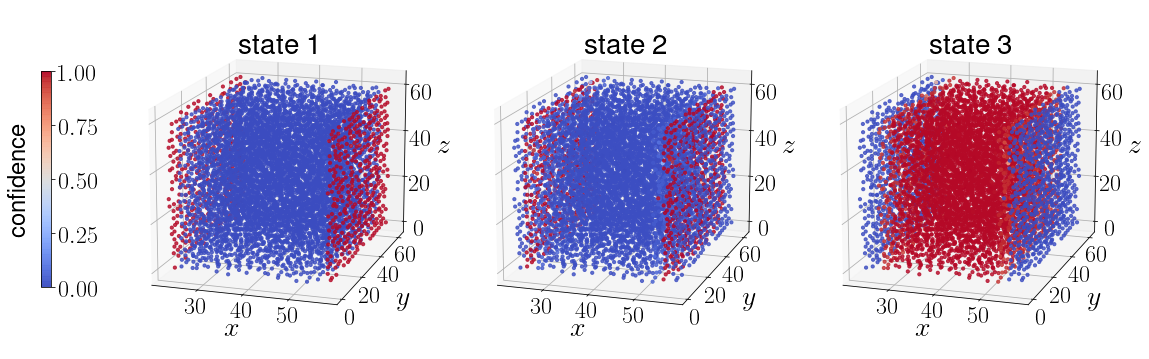}{fig:Al-distr_sub2}
    \subplot{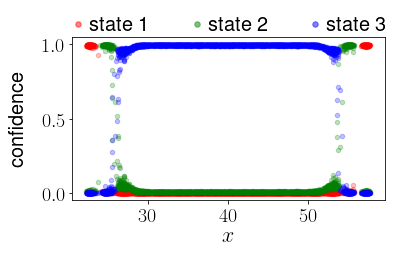}{fig:Al_prob_sub2}
    
    \caption{The distribution of states of Al atoms in Al-substrated system at 1033.5K. The substrates are omitted. (a), (c) and (e) depict the spatial distribution of probabilities of Al atoms being in different states for composition Al$100\%$Ti$0\%$, Al$95\%$Ti$5\%$ and Al$90\%$Ti$10\%$, respectively. (b), (d) and (f) are the corresponding probability distribution as a function of the z position.}
    \label{fig:Al-based}
\end{figure}

\begin{figure}[htbp]
    \def\subplotheight{0.18\textwidth}
    \def\verticalskip{0.5\baselineskip}
    \newcommand{\subplot}[2]{
        \sidesubfloat[]{
            \includegraphics[height=\subplotheight]{#1}\label{#2}
        }
    }

    \subplot{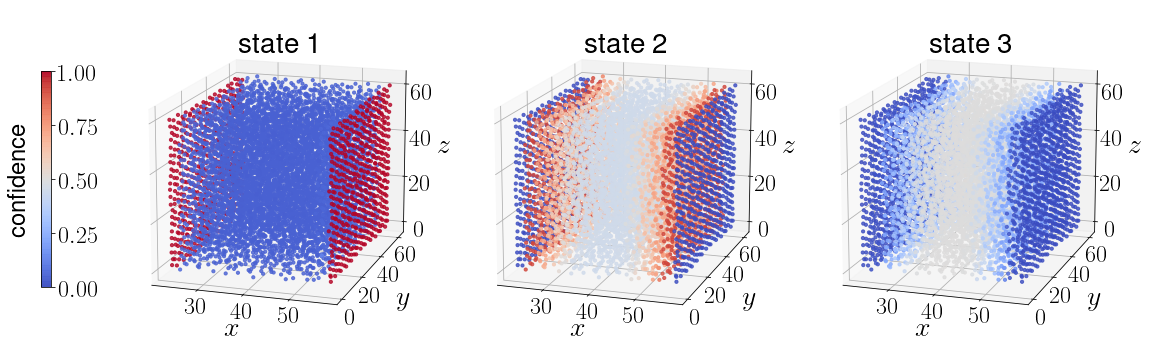}{fig:Ti-distr_sub0}
    \subplot{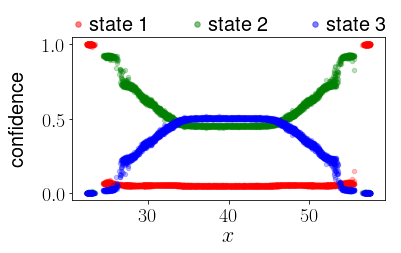}{fig:Ti_prob_sub0}
    \\

    \subplot{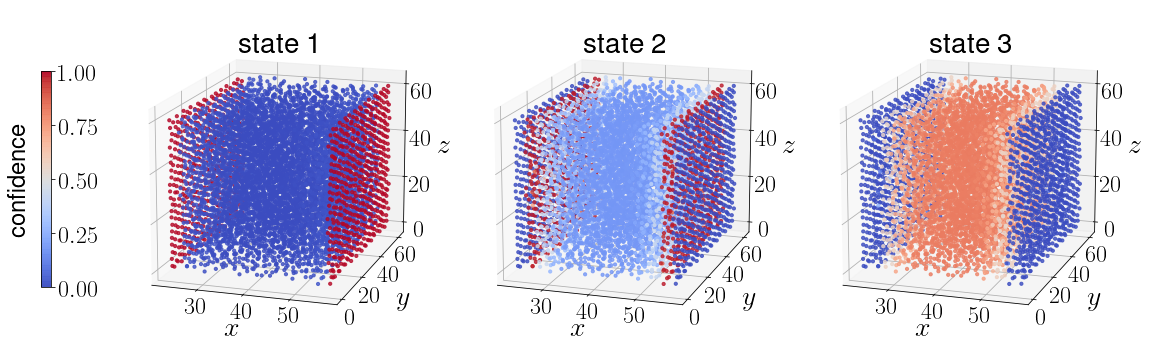}{fig:Ti-distr_sub1}
    \subplot{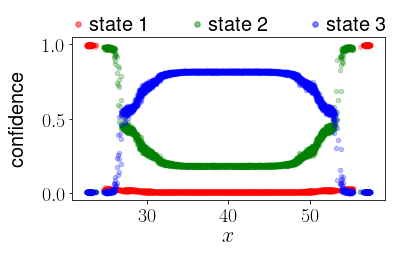}{fig:Ti_prob_sub1}
    \\

    \subplot{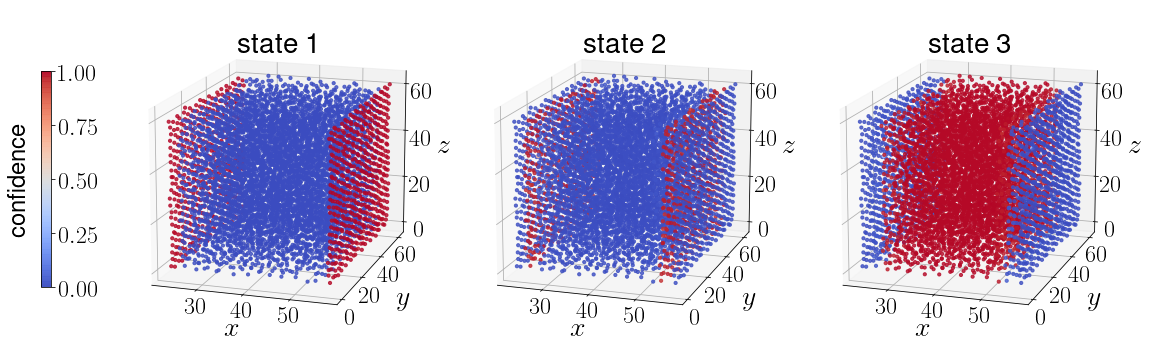}{fig:Ti-distr_sub2}
    \subplot{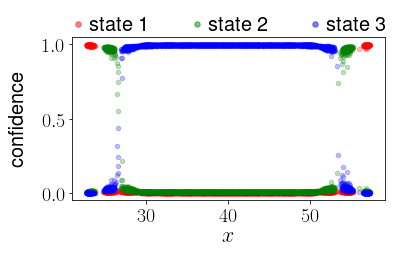}{fig:Ti_prob_sub2}
    
    \caption{The distribution of states of Al atoms in Ti-substrated system at 1033.5K. The substrates are omitted. (a), (c) and (e) depict the spatial distribution of probabilities of Al atoms being in different states for composition Al$100\%$Ti$0\%$, Al$95\%$Ti$5\%$ and Al$90\%$Ti$10\%$, respectively. (b), (d) and (f) are the corresponding probability distribution as a function of the z position.}
    \label{fig:Ti-based}
\end{figure}

The results for the Al-substrated and Ti-substrated systems at \SI{1033.5}{\kelvin} are presented in \cref{fig:Al-based} and \cref{fig:Ti-based}, respectively. In both systems, three Markov states of the dynamics of Al are observed:
\begin{itemize}
    \item state 1 (solid Al): a layer of Al atoms that are near the substrates
    \item state 2 (interfacial Al): Al atoms at the interfacial region
    \item state 3 (liquid Al): Al atoms in the mixture that is far away from the substrates
\end{itemize}
Remarkably, the MSM was trained solely on local information and not on the exact atomic positions, yet it accurately captures the spatial distribution of the learned states on a global scale.

From \cref{fig:Al-based} and \cref{fig:Ti-based} we can see that the MSMs are more confident in classifying Al atom states as the percentage of Ti in the mixture increases. For example, in Al-based systems, the MSM's confidence in identifying atoms in the middle of the mixture as liquid Al increases with the concentration of Ti in the mixture. Ti-based systems display a similar state distribution, except that in \cref{fig:Ti-based}, composition $\mathrm{Ti}0\%\mathrm{Al}100\%$, the MSM struggles to differentiate between state 2 and state 3 for many atoms. This could possibly be due to that the effect of the Ti substrate on the dynamics of Al atoms has a relatively shorter range compared to the Al substrate, so, when the composition of the mixture is as simple as pure Al, state 2 and 3 are strongly overlapping in the space.

State distributions at \SI{983.5}{\kelvin} are detailed in \cref{asec:distribution_at_983.5K}, showing similarity to the results at \SI{1033.5}{\kelvin}. Seemingly, the MSMs demonstrate greater confidence in classifying Al atom states at the lower temperature, possibly due to more distinct transition modes between states. As a supplementary analysis, we compared the distribution of Markov states with the potential energy distribution of Al atoms in \cref{asec:potential_energy_distribution}. It was noted that the two distributions do not always coincide, underlining that an understanding of the dynamics cannot be derived from the analysis of potential energy due to its inherent static nature.

\subsection{Transitions between states}
\label{sec:transitions_between_states}
\begin{figure}[htbp]
    \def\subplotheight{0.2\textwidth}
    \def\verticalskip{0.5\baselineskip}
    \newcommand{\subplot}[2]{
        \sidesubfloat[]{
            \includegraphics[height=\subplotheight]{#1}\label{#2}
        }
    }
    \subplot{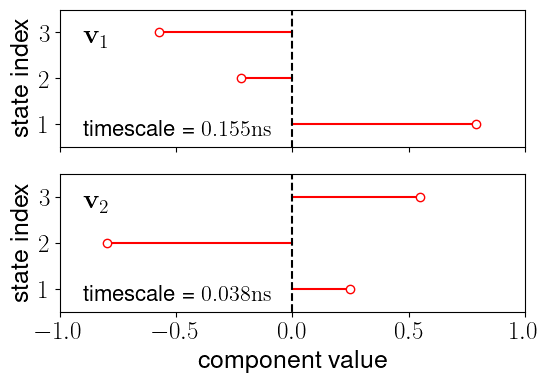}{fig:Al_eigen_0}
    \subplot{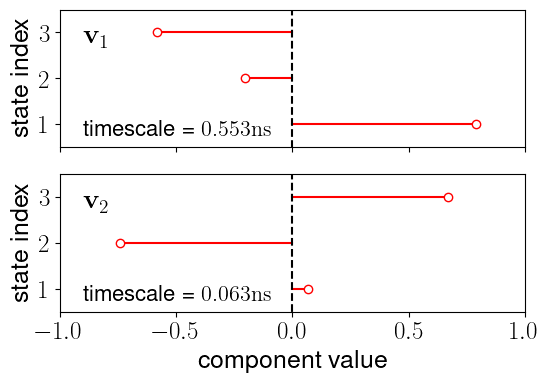}{fig:Al_eigen_2}
    \subplot{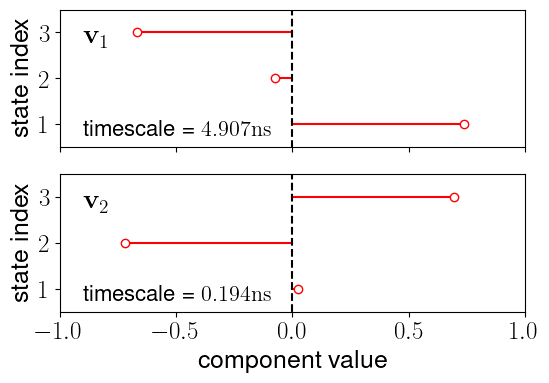}{fig:Al_eigen_3}
    
    \caption{The eigenvectors and associated timescales of Al-substrated dynamic systems with (a) Al$100\%$Ti$0\%$, (b) Al$95\%$Ti$5\%$ and (c) Al$90\%$Ti$10\%$ at 1033.5K.}
    \label{fig:Al_eigen}
\end{figure}

\begin{figure}[htbp]
    \def\subplotheight{0.2\textwidth}
    \def\verticalskip{0.5\baselineskip}
    \newcommand{\subplot}[2]{
        \sidesubfloat[]{
            \includegraphics[height=\subplotheight]{#1}\label{#2}
        }
    }
    \subplot{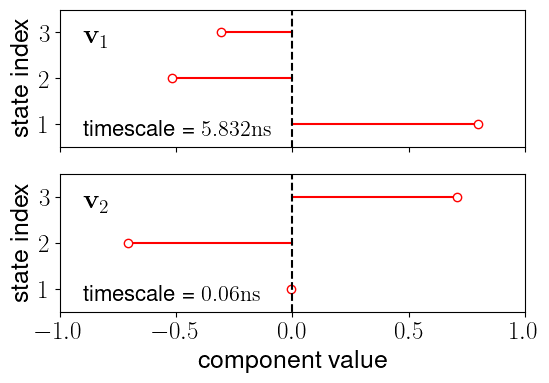}{fig:Ti_eigen_0}
    \subplot{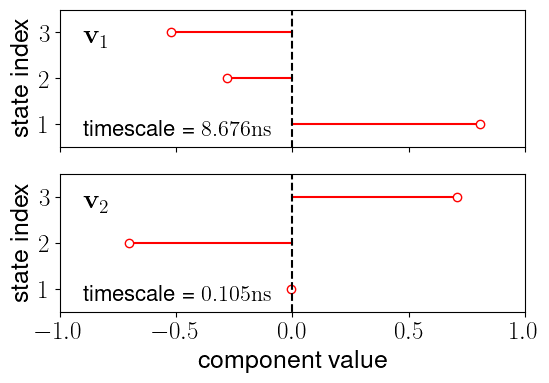}{fig:Ti_eigen_2}
    \subplot{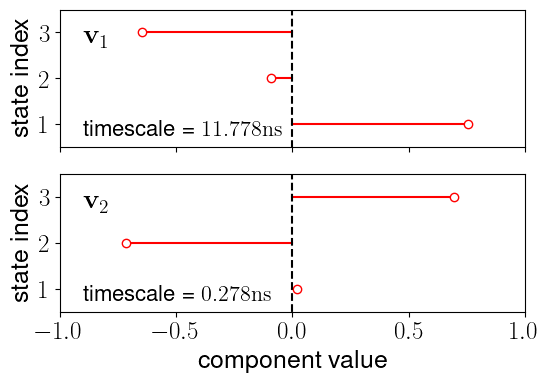}{fig:Ti_eigen_3}
    
    \caption{The eigenvectors and associated timescales of Ti-substrated dynamic systems with (a) Al$100\%$Ti$0\%$, (b) Al$95\%$Ti$5\%$ and (c) Al$90\%$Ti$10\%$ at 1033.5K.}
    \label{fig:Ti_eigen}
\end{figure}

\cref{fig:Al_eigen} and \cref{fig:Ti_eigen} display the eigenvectors obtained for Al-substrated and Ti-substrated systems with different compositions at \SI{1033.5}{\kelvin}. To understand the implication of the eigenvectors, informally, one can think of the dynamics of the system as a superposition of many ``principal modes'' with varying timescales.\cite{prinz2011probing}

The eigenvalues $\lambda_i$ determine how slow a process is. In our setup, there are three states, so the number of eigenvalues is $3$. So, in our case, there is with $1 \equiv \lambda_0 > \lambda_1 > \lambda_2$ where $\lambda_0 = 1$ corresponds to the static process. Therefore, for the purpose of analyzing the dynamic processes, we focus on $\mathbf{v}_1$ and $\mathbf{v}_2$. 

Both \cref{fig:Al_eigen} and \cref{fig:Ti_eigen}, for either Al- or Ti-substrated systems with all three compositions of the mixture, exhibit three general patterns of dynamics, as follows:
\begin{itemize}
    \item The values of $\mathbf{v}_1$ imply that the slowest process is the transition between state 1 and 3, with a small portion of the transition between state 2 and 3.
    \item The values of $\mathbf{v}_2$ imply that the second slowest process is the transition between state 2 and 3, with a small portion of the transition between state 1 and 2.
    \item Both identified processes' timescales increase dramatically as the Ti concentration in the mixture increases.
\end{itemize}
The only ``exception'' to this observation is $\mathbf{v}_1$ in \cref{fig:Ti_eigen} (a), which can be easily explained by the fact that the MSM is unable to differentiate between state 2 and 3 in this case, which is already hinted in \cref{fig:Ti-based}. Hence, the $\mathbf{v}_1$ value in \cref{fig:Ti_eigen} (a) does not contradict with the observation.


Eigenvectors and timescales for the Al and Ti-substrated systems at \SI{983.5}{\kelvin} are included in \cref{asec:timescale_at_983.5K}. It is evident from the results that the three observed dynamic patterns at \SI{1033.5}{\kelvin} persist at \SI{983.5}{\kelvin}. Furthermore, the decrease in temperature has a significant effect as the identified dynamic processes are generally two to four times slower, even with only a \SI{50}{\kelvin} decrease in temperature. For example, in the Ti-substrated system with an Al$100\%$Ti$0\%$ mixture, the timescale associated with $\mathbf{v}_1$ is \SI{5.832}{\nano\second} at \SI{1033.5}{\kelvin} and \SI{13.68}{\nano\second} at \SI{983.5}{\kelvin}. However, we note that such comparisons are not rigorous as the partitions of the dynamic states are not exactly the same even for the same system at different temperatures. Nevertheless, this analysis provides a unique perspective on understanding the properties of such dynamic systems.

These findings suggest that we do not necessarily need a system having both Ti and Al substrates to study the dynamic states. This conclusion is drawn from the observation that regardless of the substrates used, the liquid Aluminum state always exists, and can therefore be used as a consistent reference point for all processes. Consider, for instance, the Al$95\%$Ti$5\%$ mixture at 1033.5 Kelvin. The transition from solid to liquid Aluminum near an Aluminum substrate occurs on a timescale of \SI{0.553}{\nano\second}. In contrast, the same phase transition near a Titanium substrate unfolds over a notably longer timescale, approximately \SI{8.676}{\nano\second}. This indicates that in a real-world system where both substrates are present, solid Aluminum transitions to a liquid state much faster near an Aluminum substrate compared to a Titanium one. Furthermore, using substrates of identical composition simplifies the system and makes the analysis easier by creating symmetry. The results also indicate that systems with identical substrates generally have three distinct states. This implies that systems with different substrates will have at least five states -- two near the substrate states that depend on the substrate type and one liquid state in the middle. However, we need to consider that our sample size is limited to a maximum of $7500$ Aluminum atoms. Therefore, it is reasonable to use substrates of the same composition, which results in fewer states, to ensure a more robust sampling of the states.

\subsection{Implications on the diffusive dynamics}
The MD simulation results combined with the MSM analysis shed light on the mechanism of Al diffusion in the Ti--Al system. In a system with pure Ti and Al phases, with the presence of the TiAl\textsubscript{3} grain boundaries, the diffusion of Al at high temperature can be described as three stages. In the initial stage, due to the premelting effect, the Al atoms at the interfaces become liquid, and diffuse into the grain boundaries. This process widens the grain boundaries, opening up channels for Al atoms to move towards the Ti phase, which in turn could enhance the diffusion of Al atoms. In this stage, there is a net flux of Al towards Ti, whereas a very minimal amount of Ti atoms diffuse towards the Al phase. Consequently, there are Al atoms accumulating at the Ti/TiAl\textsubscript{3} interface, continuously opening the grain boundaries along the Ti surface. These are evidenced by the simulation results in \ref{sec:md_results}. Following the initial stage, with the abundance of Al atoms which has opened up the grain boundaries, the Ti atoms will also diffuse into the Al liquid, forming a liquid mixture between the Ti surface and the Al surface. In such a mixture, the transition between the solid Al near the Ti surface and the liquid Al is always slower than the transition between the solid Al near the Al surface and the liquid Al. This essentially means Al atoms in the liquid phase tend to be stably deposited onto the Ti surface more than the Al surface, enhancing the flow of Al atoms going from Al to Ti. Eventually, as the content of Ti increases in the mixture, all the dynamic processes of Al in the mixture become slower, and eventually, the liquid diffusion ceases. Results in \cref{sec:transitions_between_states} show that the timescale of the processes dramatically increase even with just a slight increase in Ti concentration in the mixture. Therefore, the diffusion processes probably will cease when the Ti content in the mixture hit some limit.

\section{Conclusion}
This study elucidates the interfacial dynamics within Ti--Al systems by combining MD simulations with MSM analyses. Our MD simulations reveal that under standard heat treatment temperatures, the interfacial Al liquefies and diffuses through TiAl\textsubscript{3} grain boundaries to the Ti/TiAl\textsubscript{3} interface, while the diffusivity of interfacial Ti remains considerably low. This results in dense accumulation of Al near the Ti/TiAl\textsubscript{3} interface, indicating that the predominant process during heat treatment is the migration of Al atoms towards this interface. This rapid migration suggests a faster formation of TiAl\textsubscript{3} at the Ti/TiAl\textsubscript{3} interface as compared to the Al/TiAl\textsubscript{3} interface. Through MSM analyses, we identified three distinct dynamic states of Al atoms: (1) Al atoms proximate to the substrate, (2) Al atoms at the interfacial region, and (3) Al atoms within the liquid mixture. By examining the eigenvectors and timescales of the dynamic processes, the slowest transition was between state 1 and 3, with the next slowest occurring between state 2 and 3. Furthermore, our findings suggest that with adequate time, as Ti diffuses gradually into the mixture, the resulting increase in Ti concentration triggers a rapid deceleration of all dynamic processes. The dynamic processes typically exhibit slower rates near the Ti substrate compared to the Al substrate. Additionally, a marginal reduction in heat treatment temperature can significantly retard all the dynamic processes. Taken together, the MD and MSM analyses provide insights into the diffusion mechanism in the Ti--Al system, characterized as a three-stage process. Initially, Al atoms diffuse into the TiAl\textsubscript{3} grain boundaries due to a premelting effect, thereby facilitating Al's migration towards the Ti phase. Subsequently, Ti atoms diffuse into the liquid Al, resulting in a mixture where Al atoms predominantly deposit onto the Ti surface. Eventually, this process terminates when the concentration of Ti in the mixture reaches a specific limit. With these insights into the diffusion mechanisms, we hope to inspire new ways to improve the control and optimization of manufacturing processes for these high-performance Ti--Al based materials to minimize defects or enhance their structural properties.

\section*{Acknowledgments}
J.Y. acknowledges support from the US National Science Foundation under awards CMMI-2038057, ITE-2236190, and EFMA-2223785, as well as the Cornell University faculty startup grant. The authors also acknowledge the computational resources provided by the NSF Advanced Cyberinfrastructure Coordination Ecosystem: Services \& Support (ACCESS) program under grant BIO210063 and the computational resources provided by the G2 cluster from Cornell University.

\section*{Competing Interests}

The Authors declare no Competing Financial or Non-Financial Interests.

\section*{Data Availability}

The datasets used and/or analysed during the current study are available from the corresponding author on reasonable request.

\section*{Author Contributions}

T.L. and J.Y. designed and conceived the research and wrote the manuscript. T.L. wrote the code to perform computations and conduct data analysis. C.T. and A.M. provided experimental data. J.Y. acquired the funding and computational resources.

\bibliographystyle{ieeetr}
\bibliography{references} 
\newpage

\section*{Appendix}
\appendix
\begin{appendices}
\renewcommand{\thefigure}{a\arabic{figure}}
\setcounter{figure}{0}

\section{Atomic environment}
\label{asec:atomic_environment}
\begin{figure}[htbp]
    \centering
    \includegraphics[width=0.3\textwidth]{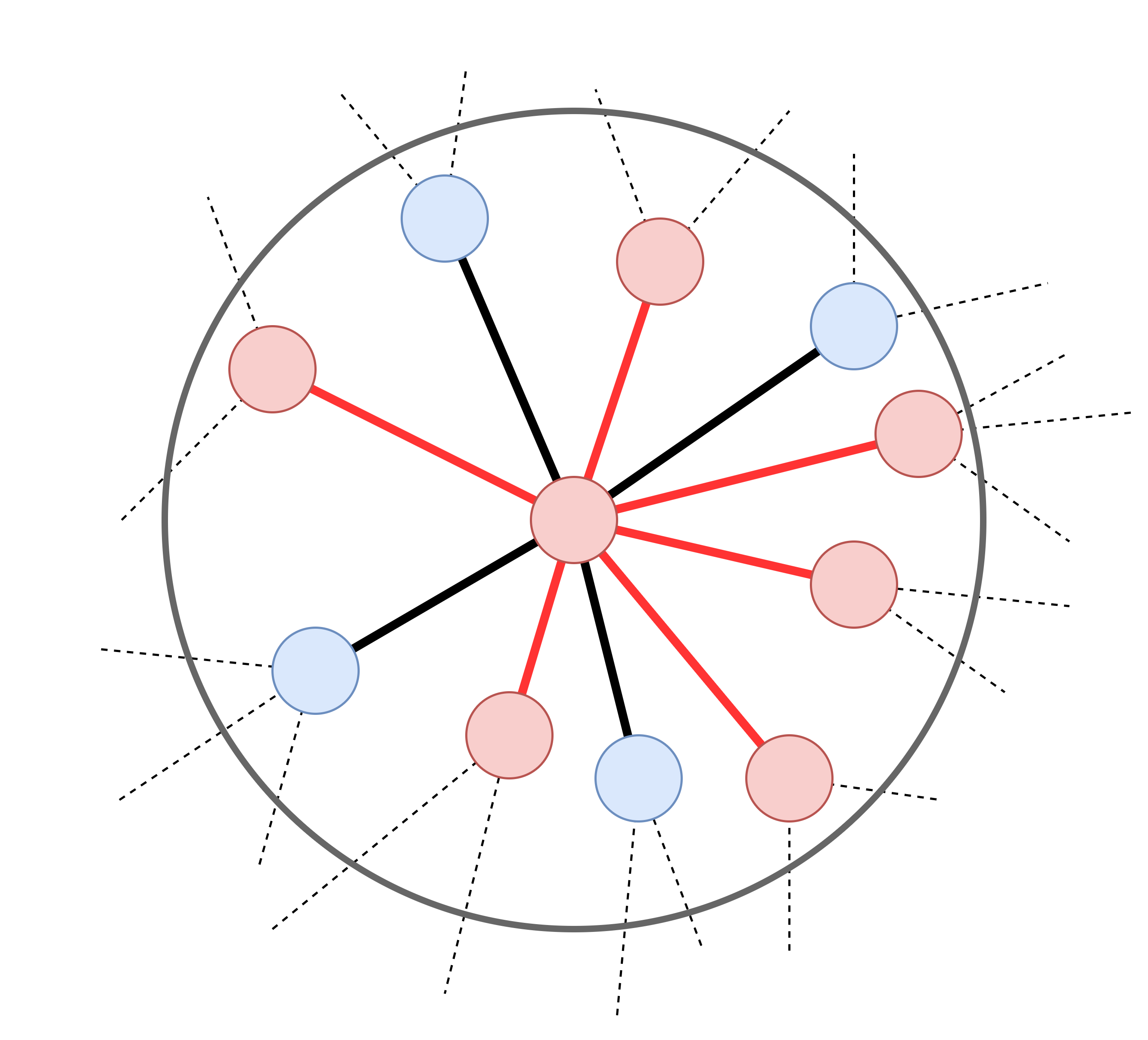}
    \caption{A schematic graph representation of the local environment of atom of interest (the middle red atom). The black edges represent the red-red bonds while the red edges represent the red-black bonds. Dashed edges represent the bonds that are not present in the local environment and the atoms outside of the local environment are omitted.}
    \label{fig:atom_graph}
\end{figure}

\cref{fig:atom_graph} shows how an atomic environment can be represented by a chemical species-colored and bond length-weighted graph.
\newpage

\section{Neural network architecture}
\label{asec:neural_network_architecture}
\begin{figure}[htbp]
    \centering
    \includegraphics[width=0.9\textwidth]{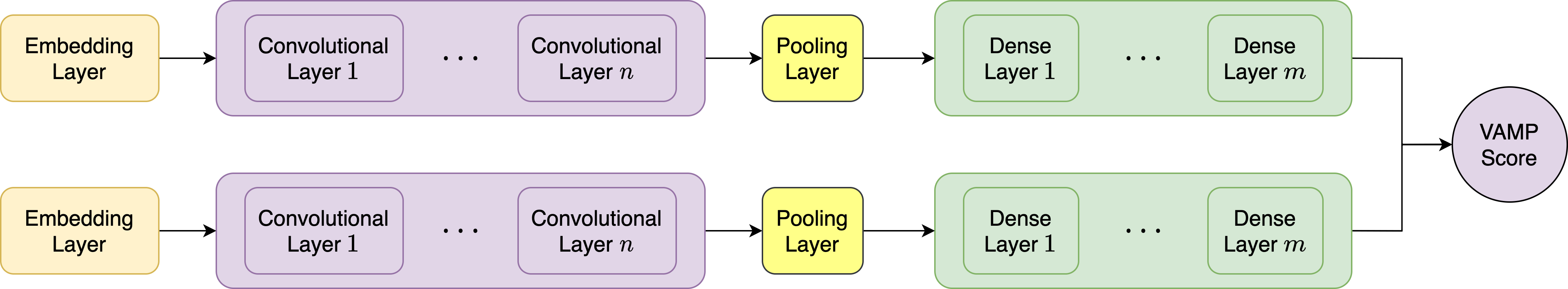}
    \caption{Structure of the GDyNets framework. The two lobes share the same network while reading different data: one reads the trajectory frame at time $t$ and the other reads at time $t+\tau$.}
    \label{fig:gdynet}
\end{figure}

The structure of the GDyNets framework is depicted in \cref{fig:gdynet}. In the context of this research, the length of the atom feature is designated as $16$. The design employs $6$ graph convolutional layers followed by $6$ dense layers succeeding the pooling layer. The training process spans $30$ epochs with a batch size set at $60$. Optimization is carried out using the default Adam optimizer. For further insights and intricate details, we recommend the reader to refer to the original paper \cite{xie2019graph}.

\newpage

\section{Distribution of states at 983.5K}
\label{asec:distribution_at_983.5K}
\begin{figure}[htbp]
    \def\subplotheight{0.18\textwidth}
    \def\verticalskip{0.5\baselineskip}
    \newcommand{\subplot}[2]{
        \sidesubfloat[]{
            \includegraphics[height=\subplotheight]{#1}\label{#2}
        }
    }

    \subplot{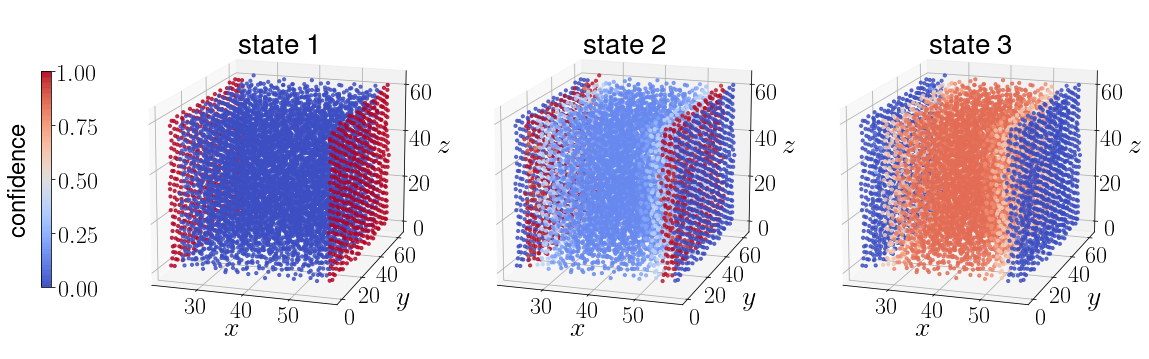}{fig:Ti_distr_sub0_983_dot_5K}
    \subplot{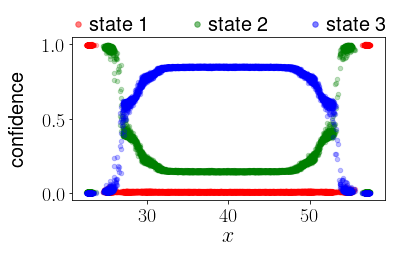}{fig:Ti_prob_sub0_983_dot_5K}
    \\

    \subplot{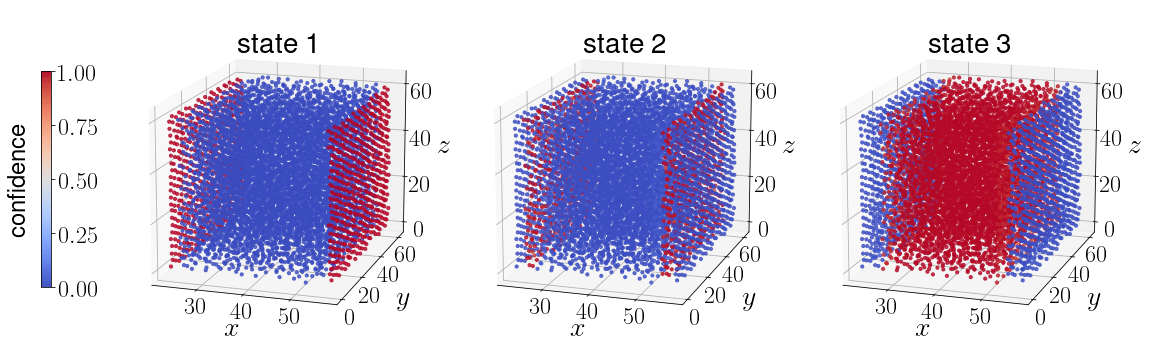}{fig:Ti_distr_sub1_983_dot_5K}
    \subplot{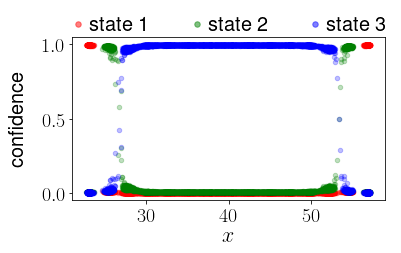}{fig:Ti_prob_sub1_983_dot_5K}
    
    \caption{The distribution of states of Al atoms in Ti-substrated system at \SI{983.5}{\kelvin}. The substrates are omitted. (a) and (c) depict the spatial distribution of probabilities of Al atoms being in different states for composition Al$100\%$Ti$0\%$ and Al$95\%$Ti$5\%$ respectively. (b) and (d) are the corresponding probability distribution as a function of the z position.}
    \label{fig:Ti_based_983_dot_5K}
\end{figure}

\begin{figure}[htbp]
    \def\subplotheight{0.18\textwidth}
    \def\verticalskip{0.5\baselineskip}
    \newcommand{\subplot}[2]{
        \sidesubfloat[]{
            \includegraphics[height=\subplotheight]{#1}\label{#2}
        }
    }

    \subplot{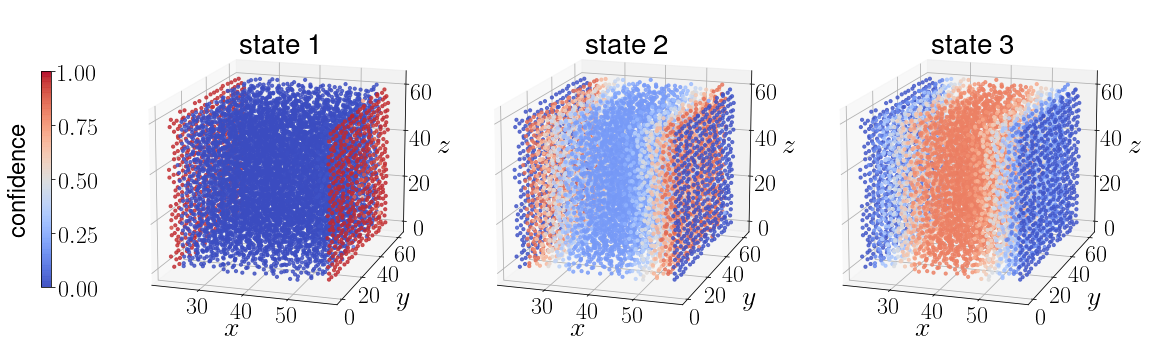}{fig:Al_distr_sub0_983_dot_5K}
    \subplot{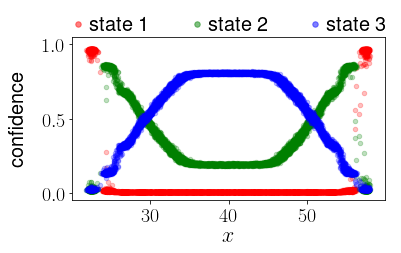}{fig:Al_prob_sub0_983_dot_5K}
    \\

    \subplot{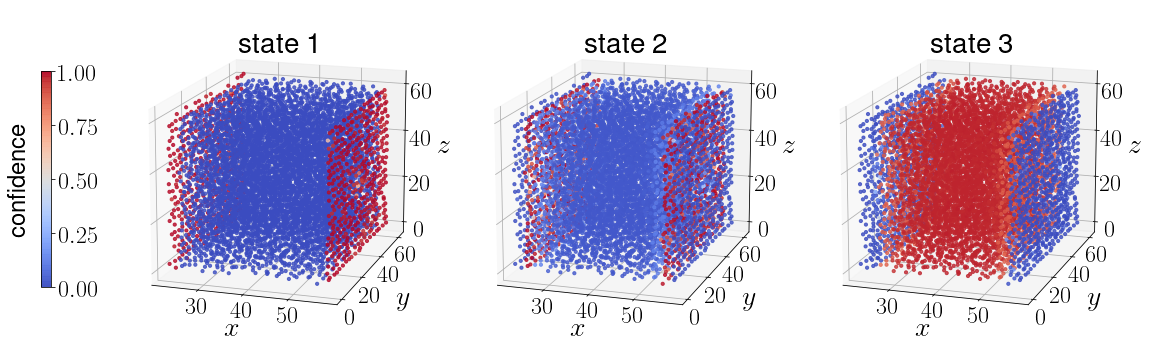}{fig:Al-distr_sub1_983_dot_5K}
    \subplot{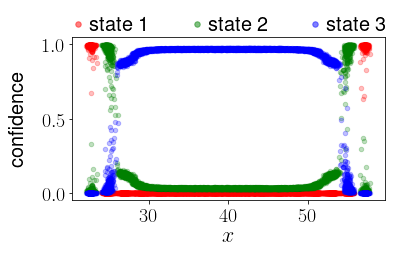}{fig:Al_prob_sub1_983_dot_5K}
    
    \caption{The distribution of states of Al atoms in Al-substrated system at \SI{983.5}{\kelvin}. The substrates are omitted. (a) and (c) depict the spatial distribution of probabilities of Al atoms being in different states for composition Al$100\%$Ti$0\%$ and Al$95\%$Ti$5\%$ respectively. (b) and (d) are the corresponding probability distribution as a function of the z position.}
    \label{fig:Al_based_983_dot_5K}
\end{figure}

The distributions of the Markov states of Al atoms at \SI{983.5}{\kelvin} are similar to the distributions at \SI{1033.5}{\kelvin}. The results are shown in \cref{fig:Ti_based_983_dot_5K} and \cref{fig:Al_based_983_dot_5K}. The Al$90\%$Ti$10\%$ composition is not included in either system at this temperature due to the slow dynamics and lack of significant dynamic behavior.

\newpage

\section{Distribution of states and distribution of potential energy}
\label{asec:potential_energy_distribution}
\begin{figure}[htbp]
    \centering
    \includegraphics[width=0.5\textwidth]{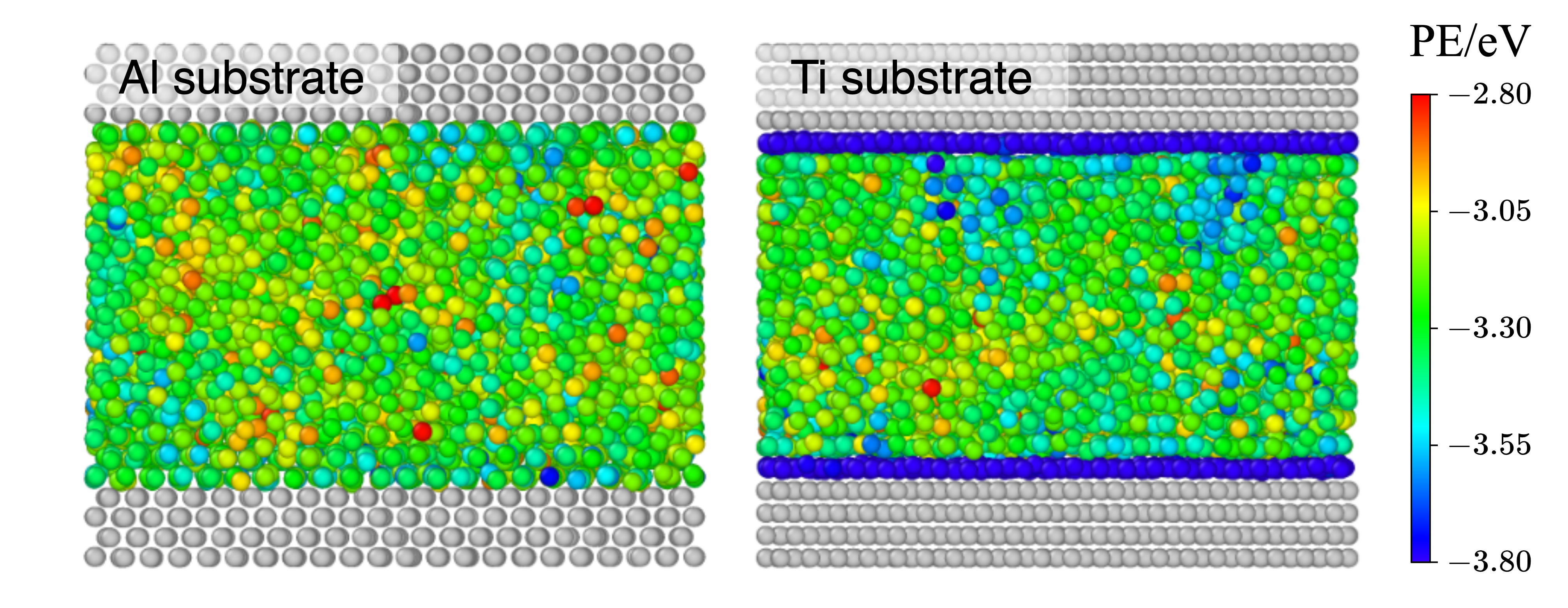}
    \caption{The atomic potential energy of Al-substrated and Ti-substrated systems with $c = \mathrm{Al}95\%\mathrm{Ti}5\%$ at \SI{1033.5}{\kelvin}.}
    \label{fig:atomic_potential}
\end{figure}

While it may be tempting to assume that the dynamics are strongly correlated with the atomic potentials, as slow transitions between states could indicate a high energy barrier, this is not always the case, as demonstrated in \cref{fig:atomic_potential}. Specifically, in the Ti-substrated system, there is a single layer of Al atoms in close proximity to the substrate that exhibits significantly lower energy than the atoms located further from the substrate. Conversely, in the Al-substrated system, such a layer does not exist. However, as discussed in \cref{sec:distribution_of_states}, the transition of the solid state Al atoms into other states (and vice versa) is slow in both systems.

\newpage
\section{Transitions between states at 983.5K}
\begin{figure}[htbp]
    \def\subplotheight{0.2\textwidth}
    \def\verticalskip{0.5\baselineskip}
    \newcommand{\subplot}[2]{
        \sidesubfloat[]{
            \includegraphics[height=\subplotheight]{#1}\label{#2}
        }
    }
    \subplot{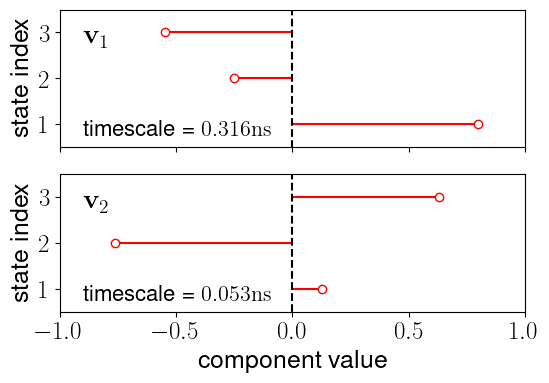}{fig:Al_eigen_0_983_dot_5K}
    \subplot{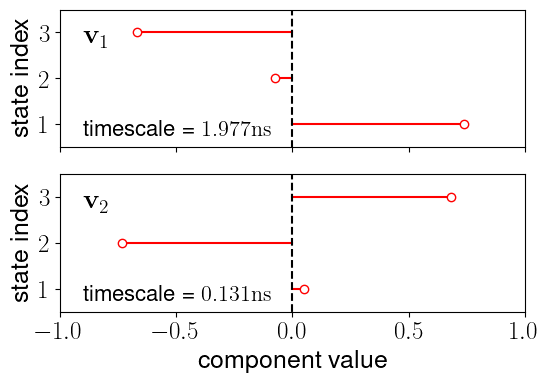}{fig:Al_eigen_1_983_dot_5K}
    
    \caption{The eigen vectors of Al-substrated dynamic systems with (a) Al$100\%$Ti$0\%$, (b) Al$95\%$Ti$5\%$ and (c) Al$90\%$Ti$10\%$ at \SI{983.5}{\kelvin}.}
    \label{fig:Al_eigen_983_dot_5K}
\end{figure}

\begin{figure}[htbp]
    \def\subplotheight{0.2\textwidth}
    \def\verticalskip{0.5\baselineskip}
    \newcommand{\subplot}[2]{
        \sidesubfloat[]{
            \includegraphics[height=\subplotheight]{#1}\label{#2}
        }
    }
    \subplot{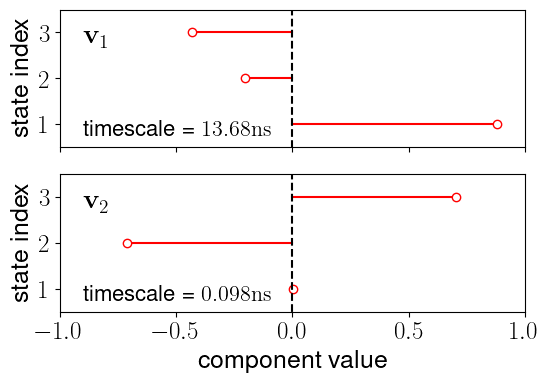}{fig:Ti_eigen_0_983_dot_5K}
    \subplot{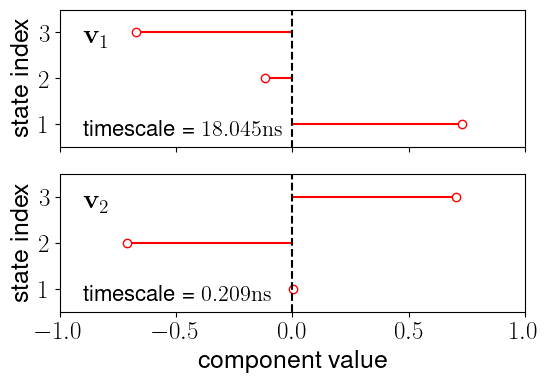}{fig:Ti_eigen_1_983_dot_5K}
    
    \caption{The eigen vectors of Ti-substrated dynamic systems with (a) Al$100\%$Ti$0\%$, (b) Al$95\%$Ti$5\%$ and (c) Al$90\%$Ti$10\%$ at \SI{983.5}{\kelvin}.}
    \label{fig:Ti_eigen_983_dot_5K}
\end{figure}

Eigenvectors and timescales for the Al and Ti-substrated systems at \SI{983.5}{\kelvin} are presented in \cref{fig:Al_eigen_983_dot_5K} and \cref{fig:Ti_eigen_983_dot_5K}, respectively. The Al$90\%$Ti$10\%$ composition is not included for the same reason stated in \cref{asec:distribution_at_983.5K}.
\label{asec:timescale_at_983.5K}

\end{appendices}

\end{document}